%% file: GMSBLesHouches.tex
\documentclass[12pt,twoside]{article}

\usepackage{epsf,cite} 
\usepackage{colordvi}

\input{SH_macros.tex}
\input{SA_macros.tex}

\begin{document}
\pagestyle{myheadings} 
\markboth
{\it Aspects of GMSB Phenomenology at TeV Colliders}
{\it Aspects of GMSB Phenomenology at TeV Colliders}

\begin{titlepage}
\thispagestyle{empty} 
\null~\vspace{-3.0cm} 
\begin{flushright} 
                           CERN-TH/2000-054 \\
hep-ph/0002191     \hfill  DESY 00/021 \\ 
\end{flushright} 
\hrule\hfill

\vspace{-0.1cm}

\hrule\hfill

\vspace{0.2cm} 
                                        
\begin{center}
{\BrickRed{\Large \bf 
Aspects of GMSB Phenomenology at TeV Colliders}} \\

\vspace{0.2cm} 

\parbox{0.75\textwidth}{\hrule\hfill}

\vspace{0.2cm}

{\Large \sl
Report of the GMSB SUSY Working Group} \\
\vspace{0.2cm} 
{\Large \sl 
Workshop ``Physics at TeV Colliders''} \\
\vspace{0.2cm}
{\Large \sl
Les Houches, 7-18 June 1999} \\ 

\vspace{0.2cm} 

\parbox{0.75\textwidth}{\hrule\hfill}

\vspace{0.5cm}

{\large {\bf S.~Ambrosanio}~$^{a,\star}$ (convener),  
        {\bf S.~Heinemeyer}~$^{b}$,
        {\bf B.~Mele}~$^{c}$,
        {\bf S.~Petrarca}~$^{c}$,
        {\bf G.~Polesello}~$^{d}$,
        {\bf A.~Rimoldi}~$^{d,e}$,
        {\bf G.~Weiglein}~$^{a}$} \\
~\\
$^a$ CERN -- {\it Theory Division,
     CH-1211 Geneva 23, Switzerland}\\
~\\
$^b$ DESY -- {\it Theory Group, 
     Notkestr. 85, D-22603 Hamburg, Germany}\\
~\\
$^c$ INFN -- {\it Sezione di Roma I e Dipartimento di Fisica, 
     Universit\`a ``La Sapienza'',}\\
     {\it p.~le Aldo Moro 2, I-00185 Roma, Italy}\\
~\\
$^d$ INFN -- {\it Sezione di Pavia e Dip.~Fis.~Nucl.~Teor.,
     Universit\`a di Pavia,} \\
     {\it via Bassi 6, I-27100 Pavia, Italy}\\
~\\
$^e$ CERN -- {\it EP Division,
     CH-1211 Geneva 23, Switzerland} \\
\end{center}

\vspace*{\fill}  

\begin{center} 
{\bf Abstract} \\
\end{center}
{\small 
The status of two on-going studies concerning important aspects of the 
phenomenology of gauge-mediated supersymmetry breaking (GMSB) 
models at TeV colliders is reported. \\
The first study deals with the characteristics of the light Higgs 
boson spectrum allowed by the (minimal and non-minimal) GMSB framework.
Today's most accurate GMSB model generation and two-loop 
Feynman-diagrammatic calculation of $\mh$ have been combined. 
The Higgs masses are shown in dependence of various model parameters
at the messenger and electroweak scales. In the minimal model, an upper
limit on $\mh$ of about 124 GeV is found for $\mt = 175$~GeV. \\
The second study is focused on the measurement of the fundamental
SUSY breaking scale $\sqrt{F}$ at the LHC in the GMSB scenario where
a stau is the next-to-lightest SUSY particle (NLSP) and decays into a 
gravitino with $c\tau_{\rm NLSP}$ in the range 0.5~m to 1~km.
This implies the measurement of mass and lifetime of long lived sleptons.
The identification is performed by determining the time of flight in the 
ATLAS muon chambers. Accessible range and precision on $\sqrt{F}$ 
achievable with a counting method are assessed.
}

\vspace*{\fill}  

\noindent 
\parbox{0.4\textwidth}{\hrule\hfill} \\ 
{\small 
$^\star$\ E-mail: {\tt ambros@mail.cern.ch} 
}  

\end{titlepage}

\setcounter{page}{0} 

\thispagestyle{empty} 
~\\ 
\newpage 
\thispagestyle{plain} 

\section{Introduction to Gauge-Mediated SUSY Breaking}
\label{sec:intro}
\noindent 
Since no superpartners have been detected at collider experiments so far,
supersymmetry (SUSY) cannot be an exact symmetry of Nature.
The requirement of ``soft'' supersymmetry breaking~\cite{StevePrimer} 
alone is not sufficient to reduce the free parameters
to a number suitable for predictive phenomenological studies. 
Hence, motivated theoretical hypotheses on the nature of SUSY breaking 
and the mechanism through which it is transmitted to the visible sector 
of the theory [here assumed to be the one predicted by the minimal SUSY 
extension of the standard model (MSSM)] are highly desirable. 
If SUSY is broken at energies of the order of the Planck mass and the SUSY 
breaking sector communicates with the MSSM sector through gravitational 
interactions only, one falls in the supergravity-inspired (SUGRA) scheme.
The most recognised alternative to SUGRA is based instead on the hypothesis
that SUSY breaking occurs at relatively low energy scales and it is mediated 
mainly by gauge interactions (GMSB)~\cite{oldGMSB,newGMSB,GR-GMSB}. 
A good theoretical reason to consider such a possibility is that it 
provides a natural, automatic suppression of the SUSY contributions 
to flavour-changing neutral current and $\cp$-violating processes. 
A pleasant consequence is that, at least in the simplest versions of GMSB, 
the MSSM spectrum and other observables depend on just a handful of 
parameters, typically

\be
M_{\rm mess}, \; N_{\rm mess}, \; \Lambda, \; \tb, \; {\rm sign}(\mu),
\label{eq:pars}
\ee

\noindent
where $M_{\rm mess}$ is the overall messenger scale; $N_{\rm mess}$ is the 
so-called messenger index, parameterising the structure of the messenger
sector; $\Lambda$ is the universal soft SUSY breaking scale felt by the
low-energy sector; $\tb$ is the ratio of the vacuum expectation 
values of the two Higgs doublets; sign($\mu$) is the ambiguity
left for the SUSY higgsino mass after conditions for correct 
electroweak symmetry breaking (EWSB) are imposed (see e.g.  
Refs.~\cite{GMSBmodels1,GMSBmodels2,AKM-LEP2,AB-LC}).

The phenomenology of GMSB (and more in general of any theory with low-energy
SUSY breaking) is characterised by the presence of a very light gravitino 
$\G$~\cite{Fayet},

\be 
m_{3/2} \equiv m_{\G} = \frac{F}{\sqrt{3}M'_P} \simeq 
\left(\frac{\sqrt{F}}{100 \; {\rm TeV}}\right)^2 2.37 \; {\rm eV},  
\label{eq:Gmass}
\ee

\noindent
where $\sqrt{F}$ is the fundamental scale of SUSY breaking, 100 TeV is a 
typical value for it, and $M'_P = 2.44 \times 10^{18}$ GeV is the reduced 
Planck mass.
Hence, the $\G$ is always the lightest SUSY particle (LSP) in these theories. 
If $R$-parity is assumed to be conserved, any produced MSSM particle will 
finally decay into the gravitino. Depending on $\sqrt{F}$, the interactions 
of the gravitino, although much weaker than gauge and Yukawa interactions, 
can still be strong enough to be of relevance for collider physics. 
As a result, in most cases the last step of any SUSY decay chain is 
the decay of the next-to-lightest SUSY particle (NLSP), which can  
occur outside or inside a typical detector or even close to the interaction 
point. The pattern of the resulting spectacular signatures is determined 
by the identity of the NLSP and its lifetime before decaying into the $\G$,

\be 
c \tau_{\rm NLSP} \simeq \frac{1}{100 {\cal B}}
\left(\frac{\sqrt{F}}{100 \; {\rm TeV}}\right)^4 
\left(\frac{m_{\rm NLSP}} {100 \; {\rm GeV}}\right)^{-5},
\label{eq:NLSPtau}
\ee 

\noindent
where ${\cal B}$ is a number of order unity depending on the nature
of the NLSP.

The identity of the NLSP [or, to be more precise, the identity of the 
sparticle(s) having a large branching ratio (BR) for decaying into the 
gravitino and the relevant SM partner] determines four main scenarios 
giving rise to qualitatively different phenomenology:

\begin{description} 

\item[Neutralino NLSP scenario:] Occurs whenever 
$m_{\NI} < (m_{\tauu} - m_{\tau})$. Here typically a decay of the $\NI$  
to $\G\gamma$ is the final step of decay chains following
any SUSY production process. As a consequence, the main inclusive signature 
at colliders is prompt or displaced photon pairs + X + missing energy. 
$\NI$ decays to $\G \Z$ and other minor channels may also be relevant 
at TeV colliders.    

\item[Stau NLSP scenario:] Defined by 
$m_{\tauu} < {\rm Min}[m_{\NI}, m_{\lR}] - m_{\tau}$, 
features $\tauu \to \G \tau$ decays, producing $\tau$ pairs or 
charged semi-stable $\tauu$ tracks or decay kinks + X + missing energy.
Here and in the following, $\ell$ stands for $e$ or $\mu$. 

\item[Slepton co-NLSP scenario:] When 
$m_{\lR} < {\rm Min}[m_{\NI}, m_{\tauu} + m_{\tau}]$, 
$\lR \to \G \ell$ decays are also open with large BR. In addition to the 
signatures of the stau NLSP scenario, one also gets $\ell^+\ell^-$
pairs or $\lR$ tracks or decay kinks. 

\item[Neutralino-stau co-NLSP scenario:] If 
$| m_{\tauu} - m_{\NI} | < m_{\tau}$ and $m_{\NI} < m_{\lR}$, 
both signatures of the neutralino NLSP and stau NLSP scenario are present
at the same time, since $\NI \leftrightarrow \tauu$ decays are not allowed
by phase space. 

\end{description}

Note that in the GMSB parameter space the relation $m_{\lR} > m_{\tauu}$ 
always holds. 
Also, one should keep in mind that the classification above is only 
valid as an indicative scheme in the limit $m_e$, $m_\mu \to 0$, neglecting 
also those cases where a fine-tuned choice of $\sqrt{F}$ and the sparticle 
masses may give rise to competition between phase-space suppressed decay 
channels from one ordinary sparticle to another and sparticle decays to the 
gravitino~\cite{AKM2}. 

In this report, we treat two important aspects of the GMSB phenomenology 
at TeV colliders:
\begin{description}
\item[(A)] The consequences of the GMSB hypotheses on the light Higgs spectrum
      using the most accurate tools available today for model generation and 
      $\mh$ calculation;
\item[(B)] Studies and possible measurements at the LHC with the ATLAS detector
      in the stau NLSP or slepton co-NLSP scenarios, with focus 
      on determining the fundamental SUSY breaking scale $\sqrt{F}$.
\end{description}

For this purpose, we generated about 30000 GMSB models under
well defined hypotheses, using the program {\tt SUSYFIRE}~\cite{SUSYFIRE}, 
as described in the following section. 

\section{GMSB Models}
\label{sec:models}

\noindent 
In the GMSB framework, the pattern of the MSSM spectrum is simple,
as all sparticle masses are generated in the same way and scale 
approximately with a single parameter $\Lambda$, which sets
the amount of soft SUSY breaking felt by the visible sector. 
As a consequence, scalar and gaugino masses are related to each
other at a high energy scale, which is not the case in other SUSY 
frameworks, e.g.\ SUGRA. Also, it is possible to impose other 
conditions at a lower scale to achieve EWSB and  
further reduce the dimension of the parameter space. 

To build our GMSB models, we adopt the usual phenomenological approach,
in particular following Ref.~\cite{AB-LC}, where problems relevant 
for GMSB physics at TeV colliders were also approached.
We do not specify the origin of the SUSY higgsino mass $\mu$, nor do we 
assume that the analogous soft SUSY breaking parameter $B\mu$ vanishes at the 
messenger scale. Instead, we impose correct EWSB to trade $\mu$
and $B\mu$ for $M_Z$ and $\tb$, leaving the sign of $\mu$ undetermined. 
However, we are aware that to build a satisfactory GMSB model one should also 
solve the latter problem in a more fundamental way, perhaps by providing a 
dynamical mechanism to generate $\mu$ and $B\mu$, possibly with values
of the same order of magnitude. This might be accomplished radiatively 
through some new interactions. However, in this case the other
soft terms in the Higgs potential, namely $m^2_{H_{1,2}}$, will be also  
affected and this will in turn change the values of $|\mu|$ and $B\mu$ 
coming from EWSB conditions~\cite{GR-GMSB,GMSBmodels1,GMSBmodels2}. 
Within the study {\bf (A)}, we are currently considering some ``non-minimal'' 
possibilities for GMSB models that to some extent take this problem into 
account, and we are trying to assess the impact on the light Higgs mass. 
We do not treat this topic here, but refer to~\cite{AHW} for further details.

To determine the MSSM spectrum and low-energy parameters, we solve 
the renormalisation group equation (RGE) evolution with the following 
boundary conditions at the $M_{\rm mess}$ scale,

\bea
M_a & = & N_{\rm mess} \Lambda 
g\left(\frac{\Lambda}{M_{\rm mess}}\right) \frac{\alpha_a}{4\pi}, 
\; \; \; (a=1, 2, 3) \nonumber \\
\tilde{m}^2 & =  & 2 N_{\rm mess} \Lambda^2 
f\left(\frac{\Lambda}{M_{\rm mess}}\right) 
\sum_a \left(\frac{\alpha_a}{4\pi}\right)^2 C_a,
\label{eq:bound}
\eea 

\noindent
respectively for the gaugino and the scalar masses. 
In Eq.~(\ref{eq:bound}), $g$ and $f$ are the one-loop and two-loop functions 
whose exact expressions can be found e.g. in Ref.~\cite{AKM-LEP2}, 
and $C_a$ are the quadratic Casimir invariants for the scalar fields.
As usual, the scalar trilinear couplings $A_f$ are assumed to vanish
at the messenger scale, as suggested by the fact that they (and not
their squares) are generated via gauge interactions with the messenger 
fields at the two loop-level only. 

 To single out the interesting region of the GMSB parameter space, 
we proceed as follows.
Barring the case where a neutralino is the NLSP and decays outside
the detector (large $\sqrt{F}$), the GMSB signatures are very spectacular
and are generally free from SM background. Keeping this in mind and being 
interested in GMSB phenomenology at future TeV colliders, we consider 
only models where the NLSP mass is larger than 100 GeV, assuming that 
searches at LEP and Tevatron, if unsuccessful, will in the end exclude 
a softer spectrum in most cases. 
We require that $M_{\rm mess} > 1.01 \Lambda$, to prevent an excess
of fine-tuning of the messenger masses, and that the mass of the 
lightest messenger scalar be at least 10 TeV. We also impose 
$M_{\rm mess} > M_{\rm GUT} \; {\rm exp}(-125/N_{\rm mess})$,
to ensure perturbativity of gauge interactions up to the 
GUT scale. Further, we do not consider models with $M_{\rm mess} \gtap 10^{5}
\Lambda$. As a result of this and other constraints, the messenger index 
$N_{\rm mess}$, which we assume to be an integer independent of the gauge 
group, cannot be larger than~8. To prevent the top Yukawa coupling from 
blowing up below the GUT scale, we require $\tb > 1.2$ (and in some 
cases $> 1.5$). This is also motivated by the current bounds from SUSY Higgs 
searches at LEP~II~\cite{mhiggsmSUGRA}.
Models with $\tb \gtap 55$ (with a mild dependence on $\Lambda$) are 
forbidden by the EWSB requirement and typically fail in giving $m_A^2 > 0$.

To calculate the NLSP lifetime relevant to our study {\bf (B)}, one needs 
to specify the value of the fundamental SUSY breaking scale $\sqrt{F}$ on a
model-by-model basis. Using perturbativity arguments, for each given 
set of GMSB parameters, it is possible to determine a lower bound
according to Ref.~\cite{AKM-LEP2}, 

\be 
\sqrt{F} \ge \sqrt{F_{\rm mess}} \equiv \sqrt{\Lambda M_{\rm mess}} > \Lambda.
\label{eq:sqrtFmin}
\ee

On the contrary, no solid arguments can be used to set an upper limit 
on $\sqrt{F}$ of relevance for collider physics, although some 
semi-qualitative cosmological arguments are sometimes evoked.

To generate our model samples using {\tt SUSYFIRE}, we used logarithmic
steps for $\Lambda$ (between about 45 TeV/$N_{\rm mess}$ and about 220 
TeV/$\sqrt{N_{\rm mess}}$, which corresponds to excluding models with
sparticle masses above $\sim 4$ TeV), $M_{\rm mess}/\Lambda$ (between
1.01 and $10^5$) and $\tb$ (between 1.2 and about 60), subject
to the constraints described above. 
{\tt SUSYFIRE} starts from the values of SM particle masses and gauge
couplings at the weak scale and then evolves up to the messenger
scale through RGE's. At the messenger scale, it imposes the boundary 
conditions (\ref{eq:bound}) for the soft sparticle masses and then 
evolves the full set of RGE's back to the weak scale. The decoupling 
of each sparticle at the proper threshold is taken into account. 
Two-loop RGE's are used for gauge couplings, third generation Yukawa
couplings and gaugino soft masses. The other RGE's are taken at the
one-loop level. At the scale $\sqrt{m_{\stopu}m_{\stopd}}$, EWSB conditions 
are imposed by means of the one-loop effective potential approach, 
including corrections from stops, sbottoms and staus. 
The program then evolves up again to $M_{\rm mess}$ and so on.
Three or four iterations are usually enough to get a good approximation
for the MSSM spectrum. 

\vspace{1.0cm} 

\hrule\hfill

\vspace{0.5cm} 

\noindent 
{\Large {\bf (A) The Light Higgs Boson Spectrum}} \\
{\Large {\bf \null~~~~~~in GMSB Models}} \\

\vspace{0.2cm} 

\hrule\hfill 

\vspace{0.2cm} 

\noindent
{\large Contribution by:} \\
{\large S.~Ambrosanio, S.~Heinemeyer, G.~Weiglein}

\vspace{0.5cm}

\section*{A.1~~Introduction}
\label{sec:A1}

\noindent
Within the MSSM, the masses of the $\cp$-even neutral Higgs bosons are 
calculable in terms of the other low-energy parameters. The mass of the
lightest Higgs boson, $\mh$, has been of particular interest, as
it is bounded to be smaller than the $\Z$~boson mass at the tree level. 
The \onel\ results~\cite{mhiggs1l,mhiggsf1l,mhiggsf1ldab,pierce} 
for $\mh$ have been supplemented in the
last years with the leading \twol\ corrections, performed in the
renormalisation group (RG)
approach~\cite{mhiggsRG1,mhiggsRG2}, in the effective
potential approach~\cite{mhiggsEP} and most recently in
the Feynman-diagrammatic (FD)
approach~\cite{mhiggsletter,mhiggslong,mhiggslle}. 
The \twol\ corrections have turned out to be sizeable. They can
lower the \onel\ results by up to 20\%.
These calculations predict an upper bound on $\mh$ of about 
$\mh \le 130$ GeV for an unconstrained MSSM with $\mt = 175$ GeV 
and a common SUSY mass scale $\msusy \le 1$ TeV.

As discussed in Sec.~\ref{sec:intro}, the GMSB scenario provides a 
relatively simple set of constraints and thus constitutes a very 
predictive and readily testable realization of the MSSM.
The main goal of the present analysis is to study the spectrum of the
lightest neutral $\cp$-even Higgs boson, $\mh$, within the GMSB framework. 
Particular emphasis is given to the maximal value of $\mh$ achievable in 
GMSB after an exhaustive scanning of the parameter space. 
Our results are discussed in terms of the GMSB constraints on the 
low-energy parameters and compared to the cases of a SUGRA-inspired or an 
unconstrained MSSM. 

\section*{A.2~~Calculation of $\mh$}
\label{sec:A2}

\noindent
To evaluate $\mh$, we employ the currently most accurate calculation 
based on the FD approach~\cite{mhiggsletter,mhiggslong,mhiggslle}. 
The most important radiative corrections to $\mh$ arise from the top and
scalar top sector of the MSSM, with the input parameters $\mt$, 
the masses of the scalar top quarks, $\mste$, $\mstz$, and the
$\Stop$-mixing angle, $\tst$. Here we adopt the conventions
introduced in Ref.~\cite{mhiggslong}.
The complete diagrammatic \onel\ result~\cite{mhiggsf1ldab} has been 
combined with the dominant \twol\ corrections of 
$\oaas$~\cite{mhiggsletter,mhiggslong} and with the subdominant 
corrections of $\ogmzmts$~\cite{mhiggsRG1,mhiggsRG2}. 
GMSB models are generated with the program {\tt SUSYFIRE}, according 
to the discussion of Sec.~\ref{sec:models}. For this study, we consider 
only models with $\tb > 1.5$~\cite{mhiggsmSUGRA} 
and $m_A > 80$ GeV~\cite{lepc}. 
In addition, we always use $\mt = 175$ GeV. A change of 1 GeV in $\mt$ 
translates roughly into a shift of 1 GeV (with the same sign) in $\mh$ 
as well. Thus, changing $\mt$ affects our results on $\mh$ in an 
easily predictable way. 

The results of the $\mh$ calculation have been implemented in the 
program \fh\ \cite{feynhiggs}. This {\tt Fortran} code has been combined 
with {\tt SUSYFIRE}, which has been used to calculate the low 
energy parameters $\mste$, $\mstz$, $\tst$, $\mu$, $M_1$, $M_2$, $\mgl$, 
$\dots$ for each of the $\sim$30000 GMSB models generated. 
These have then been passed to \fh\ for the $\mh$ evaluation in a
coherent way. Indeed, we transform the \msbar\ parameters in the 
{\tt SUSYFIRE} output into on-shell parameters before feeding
them into \fh. Compact expressions for the relevant transition 
formulas can be found in Refs.~\cite{m20,bse}.

Compared to an existing analysis in the GMSB framework~\cite{kaeding},
we use a more complete evaluation of $\mh$. This leads in particular
to smaller values of $\mh$ for a given set of input parameters in our 
analysis. Also, in Ref.~\cite{kaeding} although some GMSB scenarios with 
generalised messenger sectors were considered, the parameter space for the 
``minimal'' case with a unique, integer messenger index 
$N_{\rm mess} = N_1 = N_2 = N_3$ was not fully explored. 
Indeed, $\Lambda$ was in most cases limited to values smaller 
than 100 TeV and $M_{\rm mess}$ was fixed to $10^5$ TeV. Furthermore, 
partly as a consequence of the above assumptions, the authors did not 
consider models with $N_{\rm mess} > 4$, i.e. their requirements  
for perturbativity of the MSSM gauge couplings up to the GUT scale were 
stronger than ours. We will see in the following section that maximal 
$\mh$ values in our analysis are instead obtained for larger values of 
the messenger scale and the messenger index. 

\section*{A.3~~The Light Higgs Spectrum in GMSB}
\label{sec:A3}

\noindent 
In the following, we give some results in the form of scatter plots showing 
the pattern in GMSB for $\mh$, $m_A$ as well as other low-energy parameters
of relevance for the light Higgs spectrum. 

\begin{figure} 
\hspace{-0.5cm}
\epsfxsize=0.55\textwidth 
\epsffile{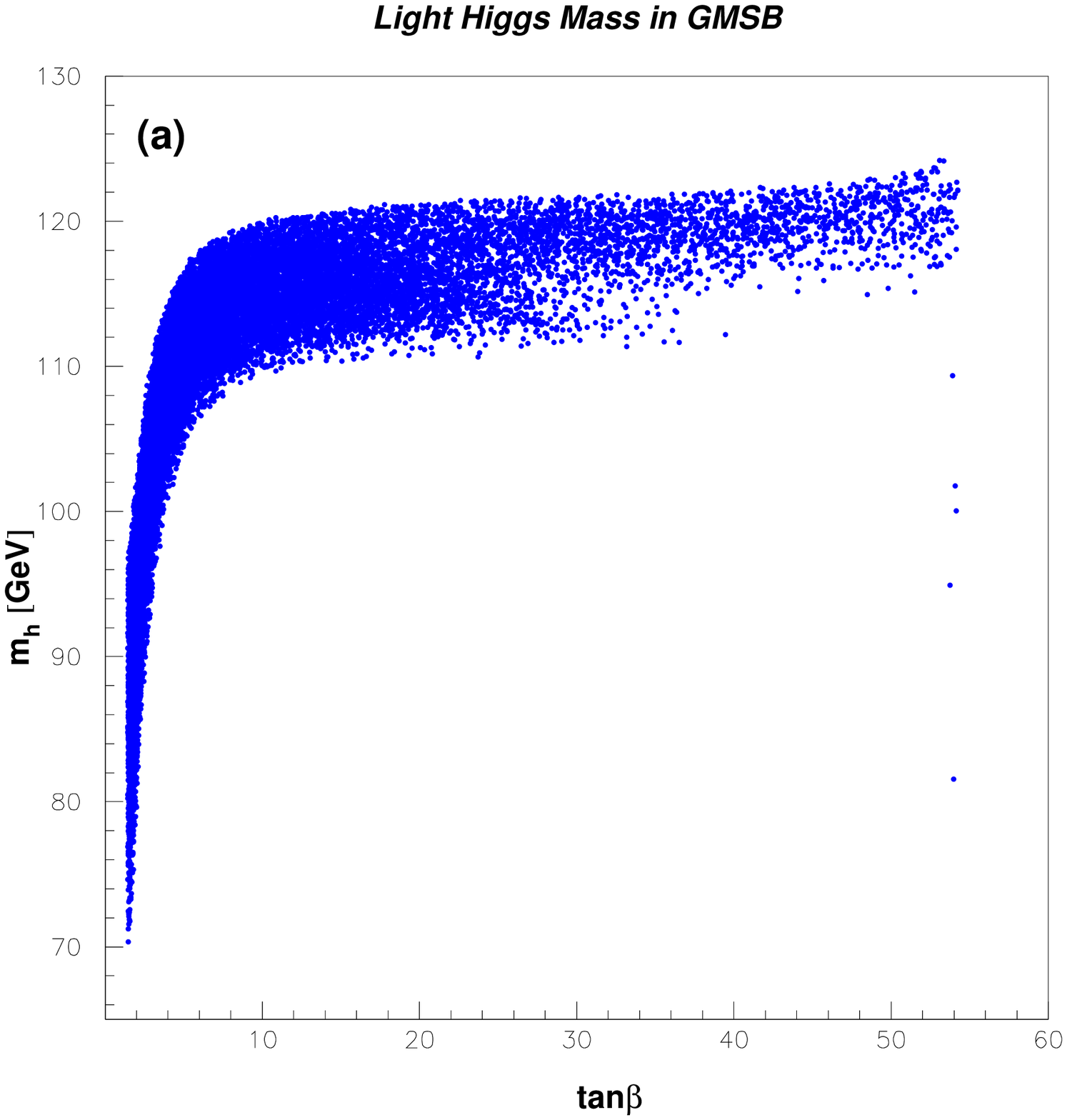}
\epsfxsize=0.55\textwidth 
\epsffile{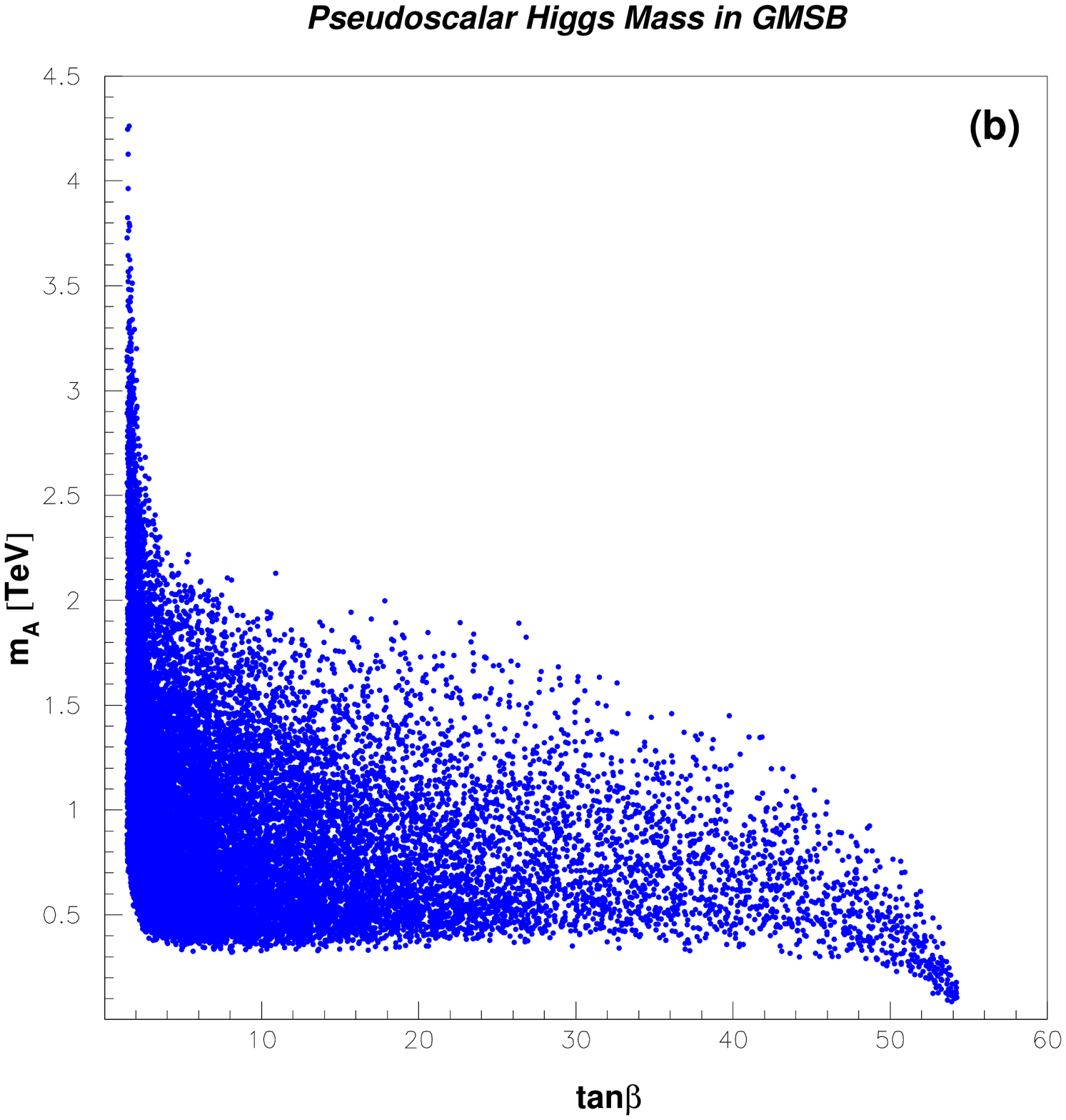}
\caption{\sl Scatter plots for the light scalar (a) and pseudoscalar (b)
Higgs masses as functions of $\tb$. Only GMSB models with 
$\tb > 1.5$, $m_A > 80$ GeV and $m_{\rm NLSP} > 100$ GeV are considered.
}
\label{fig:mh-mavstb}
\end{figure}

In Fig.~\ref{fig:mh-mavstb}(a), we show the dependence of $\mh$ on $\tb$,
where only models with $\tb > 1.5$, $m_A > 80$ GeV and
$m_{\rm NLSP} > 100$ GeV are considered, while $m_t$ is fixed to 175 GeV.
The dependence is strong for small $\tb \ltap 10$, while for larger 
$\tb$ the increase of the lightest Higgs mass is rather mild.
The maximum values for $\mh \simeq 124$ GeV are achieved for $\tb > 50$.  
It should be noted that for very large $\tb \gtap 52$, we also find a few
models with relatively small $\mh \ltap 100$ GeV. This is due
to the fact that in this case EWSB conditions tend to drive $m_A$ toward
very small values (cfr. Sec.~\ref{sec:models}). 
This is made visible by the scatter plot in Fig.~\ref{fig:mh-mavstb}(b), 
where the pseudoscalar Higgs mass is shown as a function of $\tb$. 
For such small values of $m_A$ and for large $\tb$, the relation 
$\mh \approx m_A$ holds. Thus small $\mh$ values are quite natural in 
this region of the parameter space. On the other hand, one can see that 
extremely large values of $m_A \gtap 2$ TeV can only be obtained for small 
or moderate $\tb \ltap 10$ GeV.
A comparison between Fig~\ref{fig:mh-mavstb}(a) and (b) reveals that the 
largest $\mh$ values $\gtap 123$ GeV correspond in GMSB to $m_A$ values in 
the 300--800 GeV range. 
Indeed, it has been checked that such large $\mh$ values are in general 
obtained in the FD calculation for 
$300 \ltap m_A \ltap 1000$ GeV, see Ref.~\cite{mhiggslong}.

\begin{figure} 
\hspace{3.0cm}
\epsfxsize=0.6\textwidth 
\epsffile{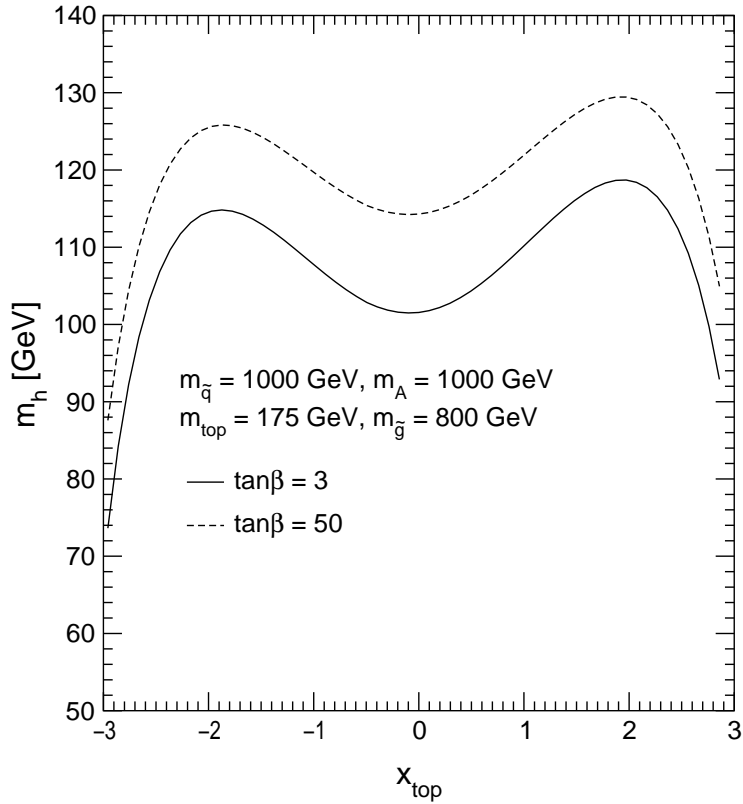}
\caption{\sl The light $\cp$-even Higgs boson mass is given as a function of 
$x_{\rm top}$ for $\tb = 3, 50$, $m_A = 1000$ GeV, a common soft SUSY 
breaking scale for the squarks, $\msq = 1000$ GeV, and a gluino 
mass $\mgl = 800$ GeV.
}
\label{fig:mhvsxt}
\end{figure}

\begin{figure} 
\hspace{-0.5cm}
\epsfxsize=0.55\textwidth 
\epsffile{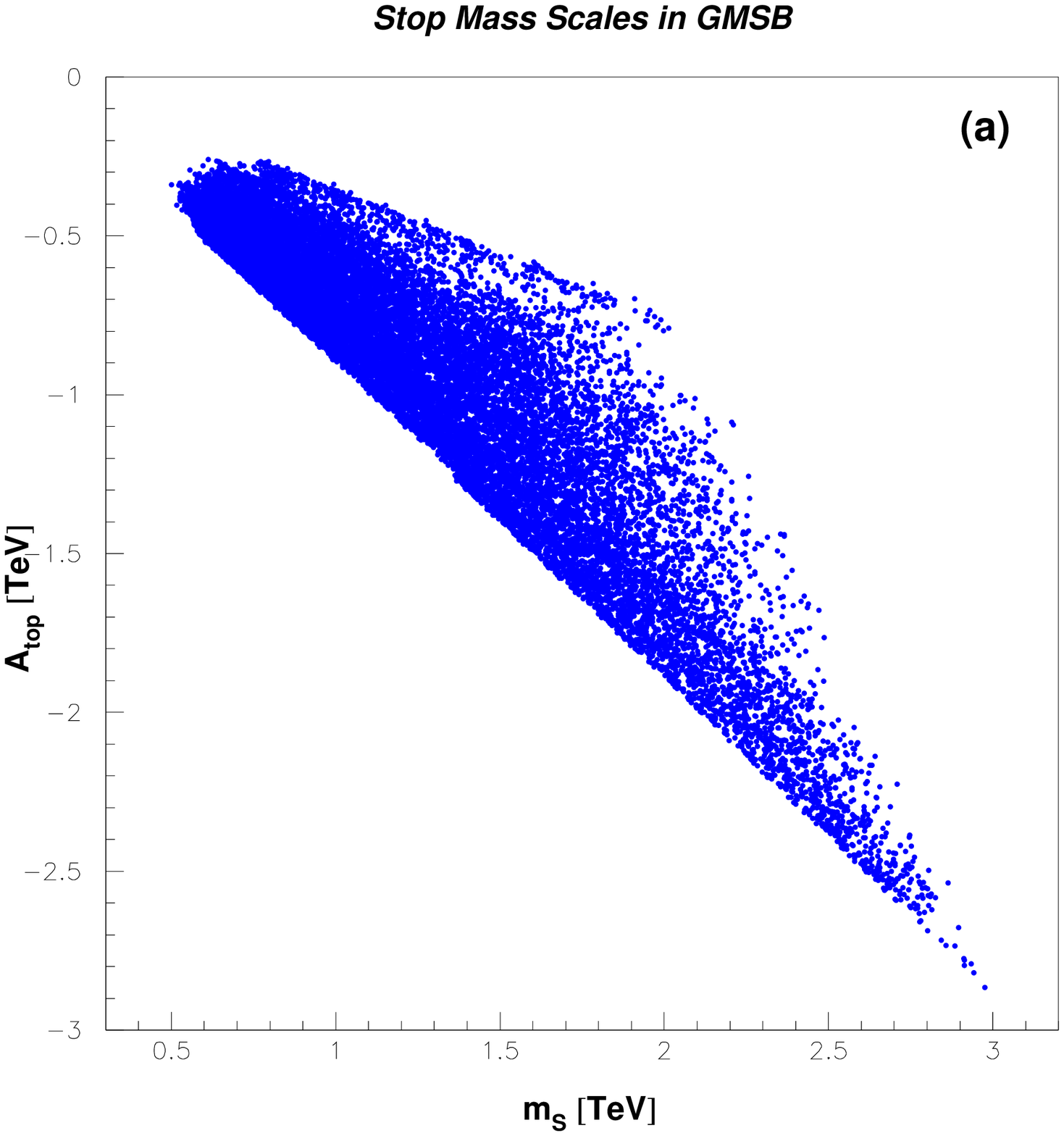}
\epsfxsize=0.55\textwidth 
\epsffile{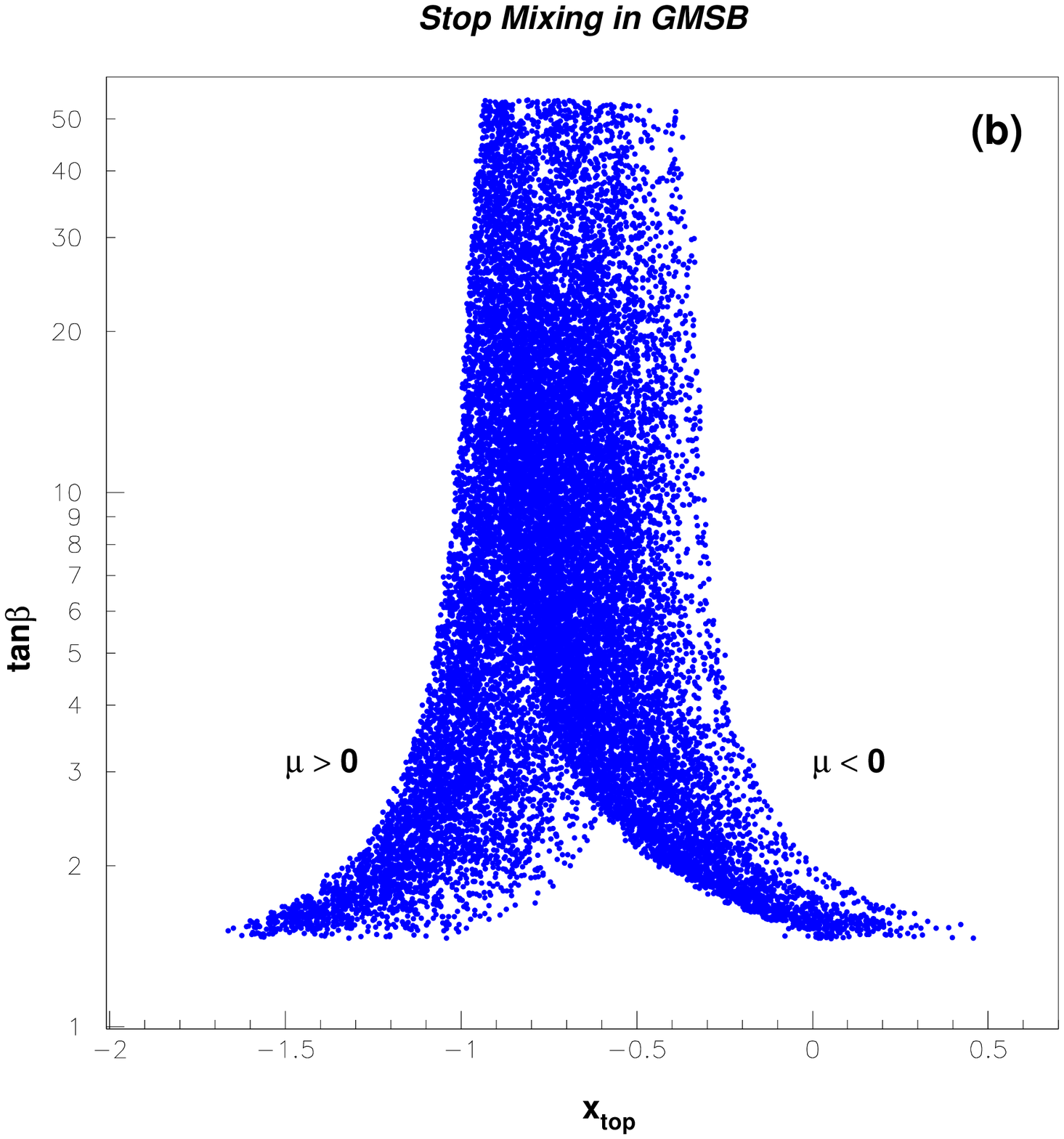}
\caption{\sl Scatter plots of $A_{\rm top}$ vs. $m_S$, the
mass scale appearing in the denominator of Eq.~(\ref{eq:stmix}) (a)
and $\tb$ vs. $x_{\rm top}$ (b).
}
\label{fig:atvsmsxtvstb}
\end{figure}

In Fig.~\ref{fig:mhvsxt}, we show the dependence of the lightest Higgs 
boson mass on the stop mixing parameter $x_{\rm top}$ defined by 

\be 
x_{\rm top} \equiv \frac{A_{\rm top} - \mu/\tb}{m_S}, \; \; \; \;
{\rm where} \; \; \; \; m_S = {\sqrt{(m_{\stopu}^2 + m_{\stopd}^2)/2}}.
\label{eq:stmix}
\ee

For equal soft SUSY breaking parameters in the stop sector with
the $D$-terms neglected, $x_{\rm top}$ corresponds to the ratio 
$X_t/M_S$ of the off-diagonal and diagonal entries in the stop mixing
matrix, see e.g. Ref.~\cite{bse}.

\begin{figure}[ht]
\hspace{-0.5cm}
\epsfxsize=0.55\textwidth 
\epsffile{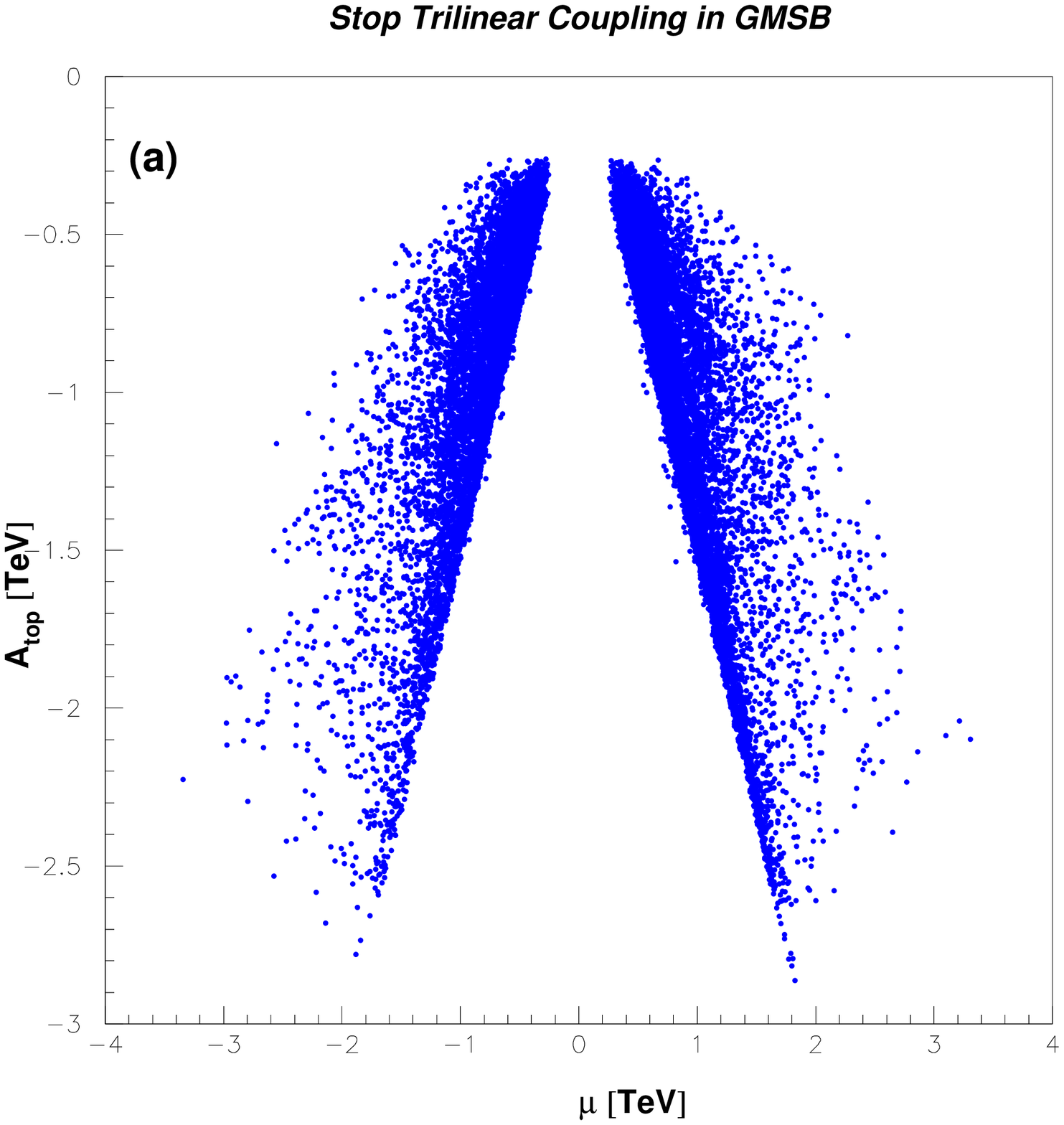}
\epsfxsize=0.55\textwidth 
\epsffile{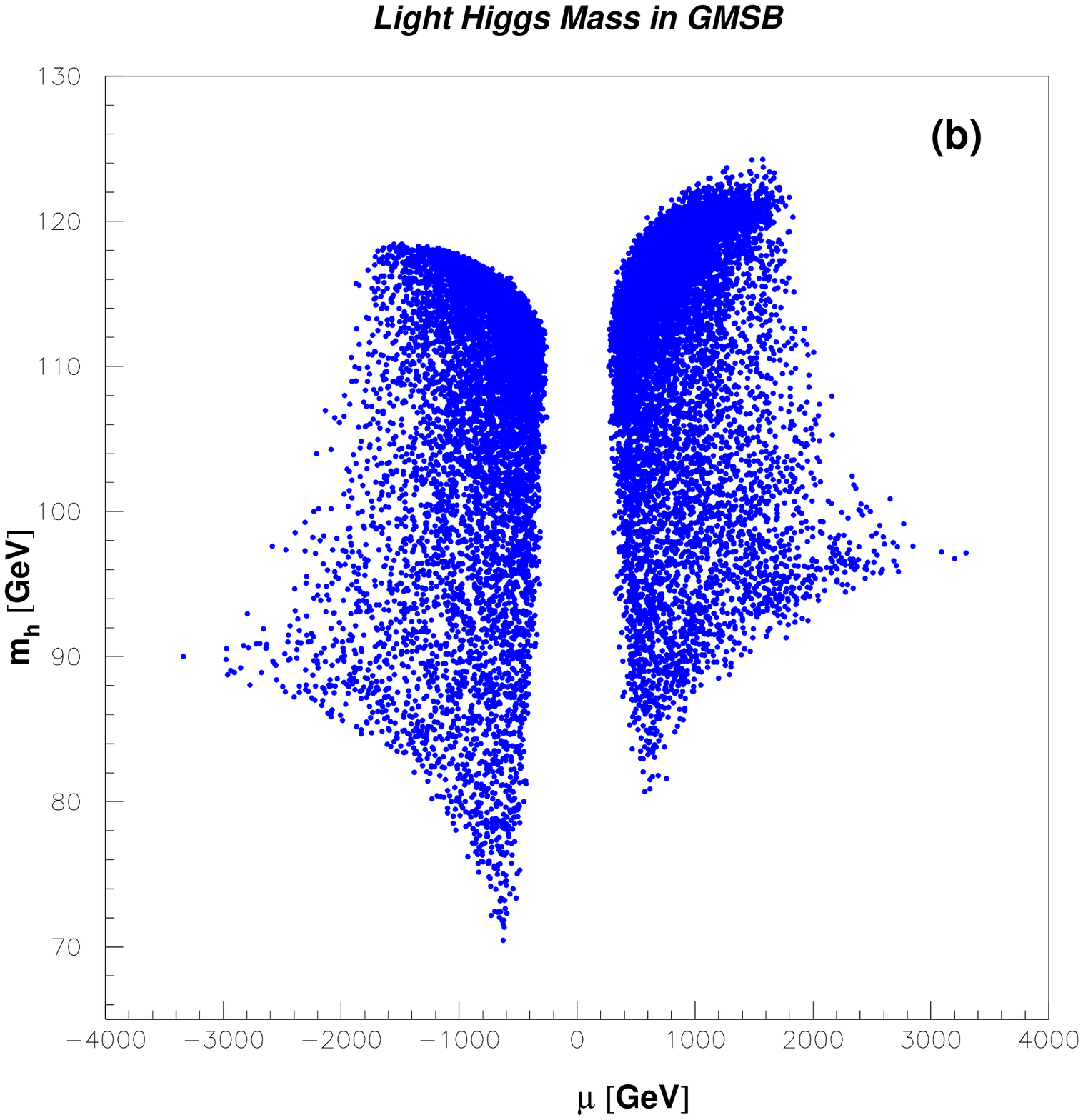}
\caption{\sl Scatter plots for $A_{\rm top}$ (a) and 
the light scalar Higgs mass (b) vs. the SUSY higgsino 
mass $\mu$ evaluated at the electroweak scale.
}
\label{fig:at-mhvsmu}
\end{figure}

Maximal $\mh$ values are obtained for $x_{\rm top} \approx \pm 2$, a
minimum is reached around $x_{\rm top} \approx 0$. Thus, for large
$\mh$ values a large numerator in Eq.~(\ref{eq:stmix}) is required.
From Fig.~\ref{fig:atvsmsxtvstb}(a), one can see that in GMSB only negative 
values of $A_{\rm top}$ are allowed at the electroweak scale, as a 
consequence of the fact that the trilinear couplings are negligible at 
the messenger scale. Due to the logarithmic dependence of $\mh$ on the stop 
masses, relatively large values of $|A_{\rm top}|$ are needed for large
$\mh$. In addition, large $\tb$ is also required. 
From Fig.~\ref{fig:atvsmsxtvstb}(b) 
one can check that this leads to values of $x_{\rm top} \approx -0.95$, 
which can only be achieved for positive $\mu$. 
Fig.~\ref{fig:at-mhvsmu}(a) shows the dependence of $A_{\rm top}$
on $\mu$. Large values of $|A_{\rm top}|$ are only reached for large 
$|\mu|$ values. Therefore maximal $h$ masses are obtained for relatively 
large and positive $\mu$, as can be seen in 
Fig.~\ref{fig:at-mhvsmu}(b).\footnote{In general, for large values of 
$|\mu|$ and $\tb$ the effects of the corrections from the $b$--$\tilde{b}$ 
sector can become important, leading to a decrease in $\mh$. For the 
GMSB models under consideration, however, this is not the case as a 
consequence of the relatively large $\tilde{b}$ masses.}

\newpage

All these arguments about the combination of low energy parameters 
needed for large $\mh$ in GMSB are summarised in Tab.~\ref{tab:one},   
where we report the 10 models in our sample that give 
rise to the highest $\mh$ values. Together with $\mh$, Tab.~\ref{tab:one} 
shows the corresponding input GMSB parameters [cfr. Eq.~(\ref{eq:pars})] 
as well as the values of the low energy parameters mentioned above.

It is interesting to note that all the models shown in Tab.~1 feature
a large messenger index and values of the messenger scale not far from 
the maximum we allowed while generating GMSB models. We could not construct
a single model with $\mh \gtap 122.5$ GeV having $N_{\rm mess} < 6$ or
$M_{\rm mess} < 10^5$ TeV, for $m_t = 175$ GeV. It is hence worth mentioning
here that our choice of imposing $M_{\rm mess}/\Lambda < 10^5 \Rightarrow 
M_{\rm mess} \ltap 2 \times 10^{10}$ GeV does not correspond to any solid 
theoretical prejudice. On the other hand it is true that 
$M_{\rm mess} \gtap 3 \times 10^8$ GeV always corresponds to gravitino masses 
larger than $\sim 1$ keV, due to Eqs.~(\ref{eq:Gmass}) and (\ref{eq:sqrtFmin}).
The latter circumstance might be disfavoured by cosmological 
arguments~\cite{cosmo}. A curious consequence is that the GMSB models with
the highest $\mh$ belong always to the stau NLSP or slepton co-NLSP scenarios.

Note also that restricting ourselves to GMSB models with 
$\Lambda < 100$ TeV, $M_{\rm mess} < 10^5$ TeV and $N_{\rm mess} \le 4$,
we find a maximal $\mh$ value of 122.2 GeV, for $\mt = 175$ GeV and 
$\tb \sim 52$. This is to be compared with the one-loop result of 
Ref.~\cite{kaeding}, $\mh$(max) = 131.7, for $\tb$ around 30 (the
assumed value of $\mt$ is not quoted).

Values for $\mh$ slightly larger than those we found here may also arise 
from non-minimal contributions to the Higgs potential, 
in connection with a dynamical generation of $\mu$ and $B\mu$ 
(cfr. Sec.~\ref{sec:models}). 
A treatment of this problem can be found in Ref.~\cite{AHW}.

\newpage

One should also keep in mind that our analysis still suffers from 
uncertainties due to unknown higher order corrections both in the RGE's
for GMSB model generation and in the evaluation of $\mh$ from low energy 
parameters. A rough estimate of these effects leads to shifts in $\mh$
not larger than 3 to 5 GeV.

\section*{A.4~~Conclusions}
\label{sec:A4}

\noindent
We conclude that in the minimal GMSB framework described above, values 
of $\mh \gtap 124.2$ GeV are not allowed for $\mt = 175$ GeV. This is 
almost 6 GeV smaller than the maximum value for $\mh$ one can achieve 
in the MSSM without any constraints or assumptions about the structure 
of the theory at high energy scales~\cite{mhiggslong,tbexcl,bench}.
On the other hand, the alternative mSUGRA framework allows values of 
$\mh$ that are $\sim 3$ GeV larger than in GMSB~\cite{mhiggsmSUGRA}. 
This makes the GMSB scenario slightly easier to explore via Higgs boson 
search. This result was expected in the light of the rather strong GMSB 
requirements, such as the presence of a unique soft SUSY breaking scale, 
the relative heaviness of the squarks and the gluino compared to 
non-strongly interacting sparticles, and the fact that the soft SUSY
breaking trilinear couplings $A_f$ get nonzero values at the electroweak 
scale only by RGE evolution.
Nevertheless, once the whole parameter space is explored, it is not true 
that mGMSB gives rise to $\mh$ values that are considerably smaller than in 
mSUGRA. Even smaller differences in the maximal $\mh$ might be present 
when considering non-minimal, complex messenger sectors~\cite{kaeding}
or additional contributions to the Higgs potential~\cite{GMSBmodels1,AHW}.
In any case, as for mSUGRA, current LEP~II or Tevatron data on Higgs boson 
searches are far from excluding mGMSB, and the upgraded Tevatron and 
the LHC will certainly be needed to deeply test any realistic SUSY model. 

\newpage 

\begin{table}[ht]
\begin{tabular}{|r||c|c|c|c|c|c|c|c|c|c|c|}  \hline
ID & $\mh$ & $N_{\rm mess}$ & $M_{\rm mess}$    & $\Lambda$ &
$\mu$ & $\tb$ & $m_A$ & $m_{\stopu}$ 
& $m_{\stopd}$ & $A_{\rm top}$ & $x_{\rm top}$ \\
& GeV &        & $10^6$ TeV    & TeV  & GeV &             
& GeV & GeV  & GeV    & GeV    &  \\ \hline \hline
A1    & 124.2 & 7 & 1.00 & 72.7 & 1470 & 53.4 & 367 & 2320
      & 2510  & -2150 & -0.90 \\ \hline 

A2    & 124.2 & 8 & 4.48 & 66.7 & 1570 & 53.1 & 436 & 2400
      & 2600  & -2310 & -0.94 \\ \hline

A3    & 123.7 & 6 & 2.07 & 87.0 & 1580 & 52.8 & 485 & 2420
      & 2630  & -2240 & -0.90 \\ \hline 

A4    & 123.7 & 8 & 4.67 & 52.4 & 1270 & 52.9 & 373 & 1930 
      & 2100  & -1850 & -0.93 \\ \hline

A5    & 123.5 & 8 & 4.89 & 51.1 & 1250 & 52.7 & 388 & 1880
      & 2050  & -1810 & -0.93 \\ \hline

A6    & 123.5 & 6 & 2.54 & 67.1 & 1260 & 53.0 & 349 & 1910 
      & 2080  & -1760 & -0.89 \\ \hline

A7    & 123.4 & 8 & 4.62 & 61.6 & 1470 & 51.9 & 549 & 2230
      & 2430  & -2160 & -0.94 \\ \hline

A8    & 123.4 & 6 & 4.15 & 88.1 & 1630 & 51.9 & 609 & 2450  
      & 2670  & -2300 & -0.91 \\ \hline
 
A9    & 123.3 & 7 & 3.77 & 70.3 & 1490 & 51.8 & 567 & 2260
      & 2460  & -2150 & -0.92 \\ \hline

A10   & 123.3 & 8 & 3.74 & 72.1 & 1677 & 50.7 & 756 & 2580 
      & 2800  & -2500 & -0.94 \\ \hline
\end{tabular}
\caption{\sl The 10 GMSB models giving rise to the highest $\mh$ values
in our sample. For each model, together with the light Higgs mass, we show 
the values of the GMSB input parameters and other low energy parameters of
interest for calculating $\mh$.}
\label{tab:one}
\end{table}

\bigskip
\subsection*{Acknowledgements}
S.~A. and S.~H. thank the organisers of the Workshop ``Physics at TeV
Colliders'', for the hospitality and the pleasant and productive atmosphere
in Les Houches. 

\newpage

\hrule\hfill

\vspace{0.5cm} 

\noindent 
{\Large {\bf (B) Measuring the SUSY Breaking Scale at the LHC}} \\
{\Large {\bf \null~~~~~~in the Slepton NLSP Scenario of GMSB Models}} \\

\vspace{0.2cm} 

\hrule\hfill

\vspace{0.2cm}

\noindent
{\large Contribution by:} \\
{\large S.~Ambrosanio, B.~Mele, S.~Petrarca, G.~Polesello, A.~Rimoldi}

\vspace{0.5cm}

\section*{B.1~~Introduction}
\label{sec:B1}

\noindent
The fundamental scale of SUSY breaking $\sqrt{F}$ is perhaps the most 
important quantity to determine from phenomenology in a SUSY theory.  
In the mSUGRA framework, the gravitino mass sets the scale of the 
soft SUSY breaking masses in the MSSM ($\sim 0.1-1$ TeV), so that $\sqrt{F}$ 
is typically large $\sim 10^{10-11}$ GeV [cfr. Eq.~(\ref{eq:Gmass})]. 
As a consequence, the interactions of the $\G$ with the other MSSM 
particles $\sim F^{-1}$ are too weak for the gravitino to be of relevance 
in collider physics and there is no direct way to access $\sqrt{F}$ 
experimentally. In GMSB theories, the situation is completely different. 
The soft SUSY breaking scale of the MSSM and the sparticle masses are set by
gauge interactions between the messenger and low energy sectors  
to be $\sim \alpha_{\rm SM}\Lambda$ [cfr. Eq.~(\ref{eq:bound})], so that 
typical $\Lambda$ values are $\sim 10-100$ TeV. On the other hand, 
$\sqrt{F}$ is subject to the lower bound (\ref{eq:sqrtFmin}) only, 
which tells us that values well below $10^{10}$ GeV and even as low as 
several tens of TeV are perfectly reasonable.
The $\G$ is in this case the LSP and its interactions are strong enough to
allow NLSP decays to the $\G$ inside a typical detector size.
The latter circumstance gives us a chance for extracting $\sqrt{F}$ 
experimentally through a measurement of the NLSP mass and lifetime
[cfr. Eq.~(\ref{eq:NLSPtau})]. 

Furthermore, the possibility of determining $\sqrt{F}$ with good 
precision opens a window on the physics of the SUSY breaking (the so-called 
``secluded'') sector and the way this SUSY breaking is transmitted to the 
messenger sector. Indeed, the characteristic scale of SUSY breaking felt 
by the messengers (and hence the MSSM sector) given by 
$\sqrt{F_{\rm mess}}$ in Eq.~(\ref{eq:sqrtFmin}) can be also determined
once the MSSM spectrum is known. By comparing the measured values 
of $\sqrt{F}$ and $\sqrt{F_{\rm mess}}$ it might well be possible to
get information on the way the secluded and messenger sector communicate
to each other. For instance, if it turns out that $\sqrt{F_{\rm mess}}
\ll \sqrt{F}$, then it is very likely that the communication occurs
radiatively and the ratio $\sqrt{F_{\rm mess}/F}$ is given by some loop 
factor. On the contrary, if the communication occurs via a direct
interaction, this ratio is just given by a Yukawa-type coupling 
constant, with values $\ltap 1$, see Refs.~\cite{GR-GMSB,AKM-LEP2}. 

An experimental method to determine $\sqrt{F}$ at a TeV scale $\epem$
collider through the measurement of the NLSP mass and lifetime was 
presented in Ref.~\cite{AB-LC}, in the neutralino NLSP scenario. 
Here, we are concerned about the same problem, but at a hadron collider,
the LHC, and in the stau NLSP or slepton co-NLSP scenarios. 
These scenarios provide a great opportunity at the LHC, since 
the characteristic signatures with semi-stable charged tracks 
are muon-like, but come from massive sleptons with a $\beta$ 
significantly smaller than 1. In particular, we perform
our simulations in the ATLAS muon detector, whose large size and 
excellent time resolution~\cite{TDR} allow a precision measurement of 
the slepton time of flight from the production vertex out to the muon 
chambers and hence of the slepton velocity. 
Moreover, in the stau NLSP or slepton co-NLSP scenarios, the 
knowledge of the NLSP mass and lifetime is sufficient to 
determine $\sqrt{F}$, since the factor ${\cal B}$ in 
Eq.~(\ref{eq:NLSPtau}) is exactly equal to 1. This is not the
case in the neutralino NLSP scenario, where ${\cal B}$ depends 
at least on the neutralino physical composition, and more information 
and measurements are needed for extracting a precise value of $\sqrt{F}$.

\section*{B.2~~Choice of the Sample Models \\ \null~~~~~~~and Event Simulation}
\label{sec:B2}

\noindent 
The two main parameters affecting the experimental measurement at the
LHC of the slepton NLSP properties are the slepton mass and momentum 
distribution. Indeed, at a hadron collider most of the NLSP's come
from squark and gluino production, followed by cascade decays. 
Thus, the momentum distribution is in general a function of the whole 
MSSM spectrum. However, one can approximately assume that most of the 
information on the NLSP momentum distribution is provided by the squark 
mass scale $m_{\tilde q}$ only (in the stau NLSP scenario or slepton 
co-NLSP scenarios of GMSB, one generally finds 
$m_{\tilde{g}} \gtap m_{\tilde q}$). 
To perform detailed simulations, we select a representative set of 
GMSB models generated by {\tt SUSYFIRE}. We limit ourselves to models 
with $m_{\rm NLSP} > 100$ GeV, motivated by the discussion in 
Sec.~\ref{sec:models}, and $m_{\tilde q} < 2$ TeV, in order to yield 
an adequate event statistics after a three-year low-luminosity run 
(corresponding to 30 fb$^{-1}$) at the LHC.
Within these ranges, we choose eight extreme points (four in the stau 
NLSP scenario and four in the slepton co-NLSP scenario) allowed by GMSB
in the ($m_{\rm NLSP}$, $m_{\tilde q}$) plane, in order to cover the 
various possibilities. 

\begin{table}
\begin{center}
\begin{tabular}{|r||r|r|r|r|c|} \hline
ID & $M_{\rm mess}$ (TeV) & $N_{\rm mess}$ 
   & $\Lambda$ (TeV)  & $\tb$ & sign($\mu$)  \\ \hline\hline
B1 & 1.79$\times 10^4$ & 3 &  26.6  &  7.22  & --   \\ \hline
B2 & 5.28$\times 10^4$ & 3 &  26.0  &  2.28  & --   \\ \hline
B3 & 4.36$\times 10^2$ & 5 &  41.9  & 53.7~  & +    \\ \hline
B4 & 1.51$\times 10^2$ & 4 &  28.3  &  1.27  & --   \\ \hline
B5 & 3.88$\times 10^4$ & 6 &  58.6  & 41.9~  & +    \\ \hline
B6 & 2.31$\times 10^5$ & 3 &  65.2  &  1.83  & --   \\ \hline
B7 & 7.57$\times 10^5$ & 3 & 104~~  &  8.54  & --   \\ \hline
B8 & 4.79$\times 10^2$ & 5 &  71.9  &  3.27  & --   \\ \hline
\end{tabular}
\caption{\sl Input parameters of the sample GMSB models chosen for our
study.}
\label{tab:tbg}
\end{center}
\end{table}

In Tab.~\ref{tab:tbg}, we list the input GMSB parameters we used to generate
these eight points, while in Tab.~\ref{tab:tbg1} we report the corresponding 
values of the stau mass, the squark mass scale and the gluino mass. 
The ``NLSP'' column indicates whether the model belongs to the stau NLSP 
or slepton co-NLSP scenario. The last column gives the total cross section 
in pb for producing any pairs of SUSY particles at the LHC. 

\begin{table}
\begin{center}
\begin{tabular}{|r||r|c|r|r|c|} \hline
ID &  $m_{\tilde\tau_1}$ (GeV) 
   & ``NLSP'' & $m_{\tilde q}$ (GeV) & $m_{\tilde g}$ (GeV) 
   & $\sigma$ (pb) \\ \hline \hline
B1 &  100.1 & $\tilde\tau$  &  577 &  631 & 42~~~~~ \\ \hline
B2 &  100.4 & $\tilde\ell$  &  563 &  617 & 50~~~~~ \\ \hline
B3 &  101.0 & $\tilde\tau$  & 1190 & 1480 & ~0.59~  \\ \hline
B4 &  103.4 & $\tilde\ell$  &  721 &  859 & 10~~~~~ \\ \hline
B5 &  251.2 & $\tilde\tau$  & 1910 & 2370 & ~0.023  \\ \hline
B6 &  245.3 & $\tilde\ell$  & 1290 & 1410 & ~0.36~  \\ \hline
B7 &  399.2 & $\tilde\tau$  & 2000 & 2170 & ~0.017  \\ \hline
B8 &  302.9 & $\tilde\ell$  & 1960 & 2430 & ~0.022  \\ \hline 
\end{tabular}
\caption{\sl Features of the sample GMSB model points studied.}
\label{tab:tbg1}
\end{center}
\end{table}

For each model, the events were generated with the {\tt ISAJET} 
Monte Carlo \cite{isajet} that incorporates the calculation of 
the SUSY mass spectrum and branching fraction 
using the GMSB parameters as input. We have checked that for 
the eight model points considered the sparticle masses 
calculated with {\tt ISAJET} are in good agreement with the output  
of {\tt SUSYFIRE}.

The generated events were then passed through {\tt ATLFAST} \cite{ATLFAST}, 
a fast particle-level simulation of the ATLAS detector. The {\tt ATLFAST}
package, however, was only used to evaluate the efficiency of the calorimetric
trigger that selects the GMSB events. The detailed response of the detector 
to the slepton NLSP has been parametrised for this work using the results  
of a full simulation study, as described in the next section.

\section*{B.3~~Slepton detection}
\label{sec:B3} 

The experimental signatures of heavy long-lived charged 
particles at a hadron collider have already been studied both 
in the framework of GMSB and in more general scenarios 
\cite{leandro,drtata, femoroi,marthom}.
The two main observables one can use to separate
these particles from muons are the high specific ionisation
and the time of flight in the detector.

We concentrate here on the measurement of the time of flight, 
made possible by the timing precision ($\ltap 1$~ns) 
and the size of the ATLAS muon spectrometer.

It was demonstrated with a full simulation of the ATLAS muon 
detector~\cite{gpar} that the $\beta$ of a particle can be measured 
with a resolution that can be approximately parameterised as 
$\sigma(\beta)/\beta^2=0.028$.
The resolution on the transverse momentum measurement for heavy
particles is found to be comparable to the one expected for muons.
We have therefore simulated the detector response to NLSP sleptons 
by smearing the slepton momentum and $\beta$ according 
to the parameterisations in Ref.~\cite{gpar}. 

An important issue is the online selection of the SUSY events.
We have not made any attempt to evaluate whether the heavy sleptons 
can be selected using the muon trigger. For the event selection, we 
rely on the calorimetric $E_{T}^{\rm miss}$ trigger, 
consisting in the requirement of at least a hadronic jet with $p_T>50$~GeV,
and a transverse momentum imbalance calculated only from the energy 
deposition in the calorimeter larger than 50~GeV. We checked that 
this trigger has an efficiency in excess of 80\% for all the considered
models. 

A detailed discussion of the experimental assumptions underlying
the results presented here is given in Ref.~\cite{AMPPR}.

\section*{B.4~~Event Selection and Slepton Mass Measurement} 
\label{sec:B4} 

\noindent
In order to select a clean sample of sleptons, we apply the
following requirements: 

\begin{itemize}
\item
at least a hadronic jet with $P_T>50$~GeV and a calorimetric \\
\mbox{$E_T^{\rm miss}>50$~GeV} (trigger requirement);
\item
at least one candidate slepton satisfying the following cuts: 
\begin{itemize}
\item
$|\eta|<$2.4 to ensure that the particle is in the
acceptance of the muon trigger chamber, and therefore 
both coordinates can be measured;
\item
$\beta_{\rm meas}<0.91$, where $\beta_{\rm meas}$ is the $\beta$ of the
particle measured with the time of flight in the precision chambers;
\item
The $P_T$ of the slepton candidate, after the energy loss in the
calorimeters has been taken into account, must be larger than
10~GeV, to ensure that the particle traverse all of the muon stations.
\end{itemize}
\end{itemize} 

Considering an integrated luminosity of 30~fb$^{-1}$, 
a number of events ranging from a few hundred for the models with
2~TeV squark mass scale to a few hundred thousand for a 500~GeV
mass scale survive these cuts and can be used for measuring the 
NLSP properties.

From the measurements of the slepton momentum and of particle $\beta$,
the mass can be determined using the standard relation 
$m=p\frac{\sqrt{1-\beta^2}}{\beta}$. 
For each value of $\beta$ and momentum, the measurement error is
known and it is given by the parametrisations in Ref.~\cite{gpar}.
Therefore, the most straightforward way of measuring the mass 
is just to use the weighted average of all the masses 
calculated with the above formula. 

In order to perform this calculation, the particle momentum is needed,
which implies measuring the $\eta$ coordinate. In fact, with the precision 
chambers only one can only measure the momentum components  
transverse to the beam axis. 

The measurement of the second coordinate must be provided by the 
trigger chambers, for which only a limited time window around the beam 
crossing is read out, therefore restricting the $\beta$ range where 
this measurement is available. Hence, we have evaluated the achieved 
measurement precision for two different $\beta$ intervals: $0.6<\beta<0.91$ 
and $0.8<\beta<0.91$ for the eight sample points.  
We found a statistical error well below the 0.1\% level for those model
points having $m_{\tilde q} < 1300$ GeV. Even for the three models 
(B5, B7, B8) with lower statistics ($m_{\tilde q} \simeq 2$ TeV), the 
error stays below the 0.4\% level. 

Many more details, tables and figures about this part of our study 
can be found in Ref.~\cite{AMPPR}. 

\section*{B.5~~Slepton Lifetime Measurement}
\label{sec:B5}

\noindent 
The measurement of the NLSP lifetime at a high energy $e^+e^-$ collider
was studied in detail in Ref.~\cite{AB-LC} for the neutralino NLSP case.
Similar to that study, the measurement of the slepton NLSP lifetime we are 
interested in here can be performed by exploiting the fact that two NLSP's 
are produced in each event. 
One can therefore select $N_1$ events where a slepton is detected through 
the time-of-flight measurement described above, count the number of 
times $N_2$ when a second slepton is observed and use this information  
to measure the lifetime. 
Although in principle very simple, in practice this method requires 
an excellent control on all possible sources of inefficiency for 
detecting the second slepton.

\begin{figure} 
\hspace{3.0cm}
\epsfxsize=0.7\textwidth
\epsffile{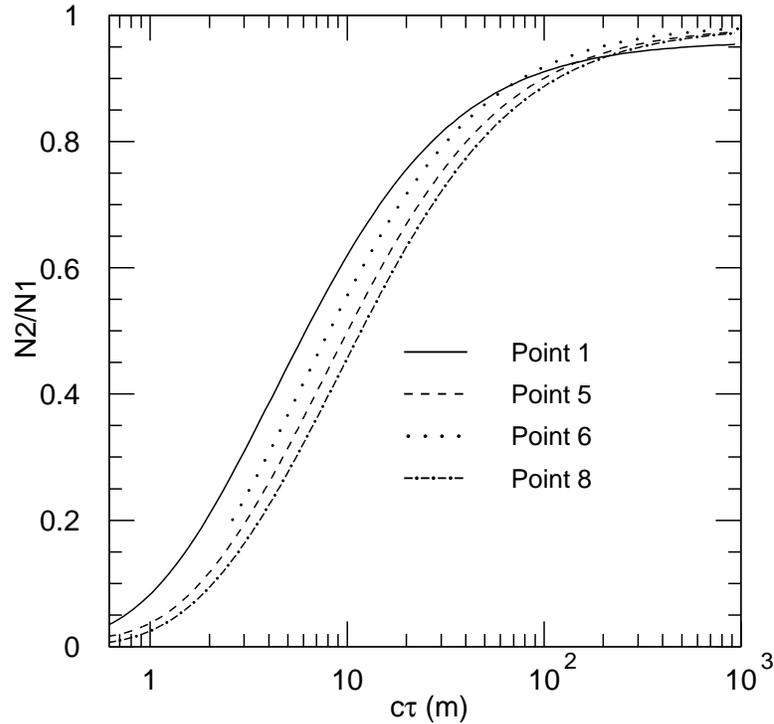}
\caption{\sl The ratio $R=N_2/N_1$ defined in the text as a function 
of the slepton lifetime $c\tau$. Only the curves corresponding to 
the model points B1, B5, B6, B8 are shown.} 
\label{fig:ratio}
\end{figure} 

We give here the basis of the method, without mentioning the experimental 
details. We provide an estimate of the achievable statistical error for 
the models considered and a parametrisation of the effect on the lifetime
measurement of a generic systematic uncertainty on the slepton efficiency.
In case the sparticle spectrum and BR's can be measured from the SUSY events, 
as e.g. shown in Ref.~\cite{ihfp}, an accurate simulation of all the SUSY 
production processes can be performed, and the results from this section are 
representative of the measurement precision achievable in a real experiment.

Another method based on the same principles, but assuming minimal knowledge 
of the SUSY spectrum, is described in Ref.~\cite{AMPPR}, where 
a detailed estimate of the achievable systematic precision is given.

We define $N_1$ starting from the event sample 
defined by the cuts discussed in Sec.~\ref{sec:B4}, 
with the additional requirement that, 
for a given value of the slepton lifetime, 
at least one of the produced sleptons decays at 
at a distance from the interaction vertex $>10$~m,
and is therefore reconstructed in the muon system.
For the events thus selected, we define $N_2$ as the 
subsample where a second particle with a transverse momentum 
$>10$~GeV is identified in the muon system. The search for the
second particle should be as inclusive as possible, in order to minimise
the corrections to the ratio. In particular, the cut $\beta_{\rm meas}<0.91$ 
is not applied, but  particles with a mass measured from $\beta$ and momentum 
incompatible with the measured slepton mass are rejected. This leaves
a background of high momentum muons in the sample that can be statistically
subtracted using the momentum distribution of electrons. 
The ratio

\be
R=\frac{N_2}{N_1}
\label{eq:ratio}
\ee

\noindent
is a function of the slepton lifetime. Its dependence on the
NLSP lifetime $c\tau$ in metres in shown in Fig.~\ref{fig:ratio}
for four among our eight sample models. The curves for the model
points not shown are either very similar to one of the curves we 
show or are mostly included between the external curves corresponding to 
points B1 and B8, thus providing no essential additional information.
Note that the curve for model 6 starts from $c\tau = 2.5$~m and not
from $c\tau = 50$~cm, as for the other models. This is due to the 
large value of $M_{\rm mess}$ (cfr. Tab.~\ref{tab:tbg}), determining
a minimum NLSP lifetime allowed by theory which is macroscopic in 
this case [cfr. Eqs.~(\ref{eq:NLSPtau}) and (\ref{eq:sqrtFmin})].

\begin{figure}[ht] 
\epsfxsize = \textwidth 
\epsffile{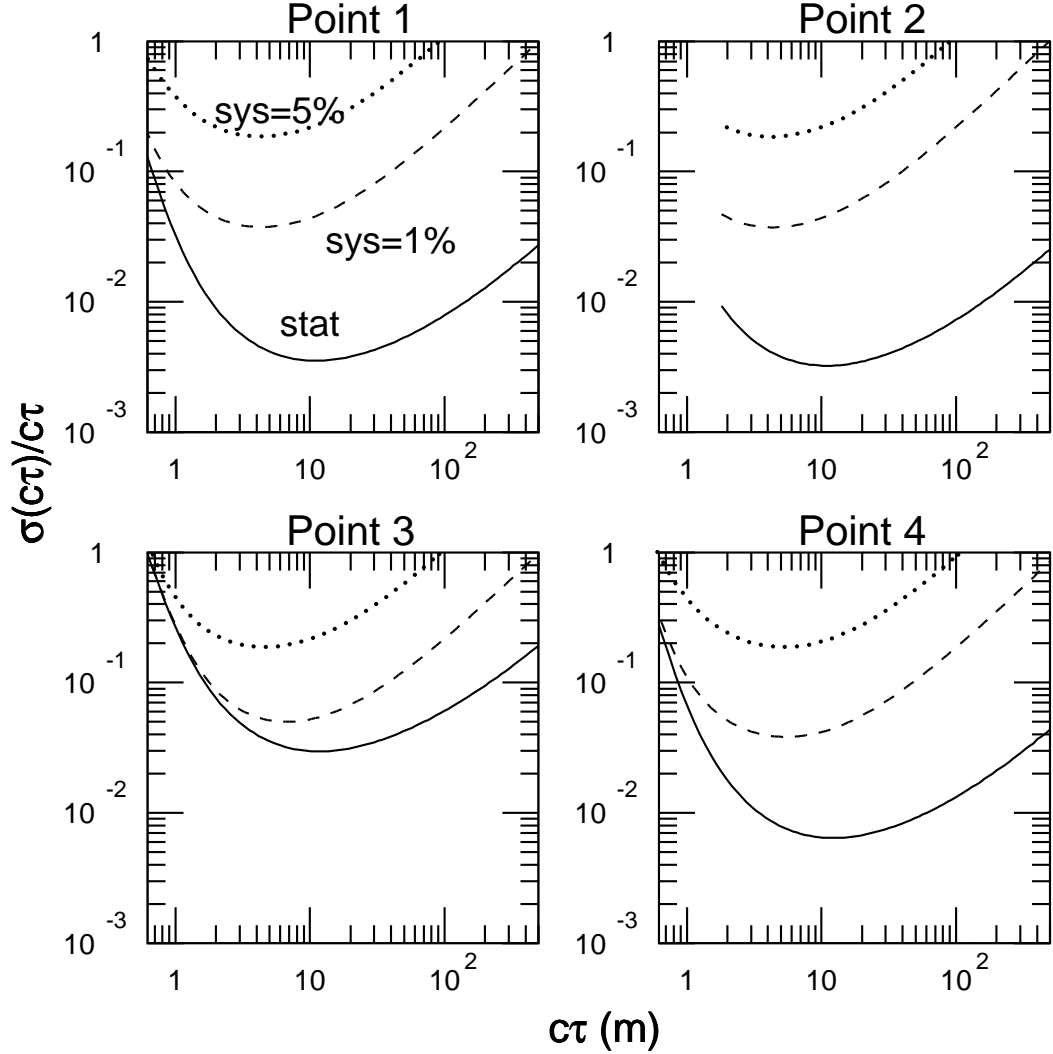}
\caption{\sl Fractional error on the measurement of the 
slepton lifetime $c\tau$, for model sample points B1 to B4.
We assume an integrated luminosity of 30~fb$^{-1}$. 
The curves are shown for three different assumptions on the 
fractional systematic error on the $R$ measurement: 
statistical error only (full line), 1\% systematic error (dashed line),
5\% systematic error (dotted line).
}
\label{fig:sigctau1}
\end{figure}

The probability for a particle of mass $m$, momentum $p$ and proper 
lifetime $\tau$ to travel a distance $L$ before decaying is
given by the expression

\be
P(L)=e^{-mL/pc\tau}.
\label{eq:prob} 
\ee

$N_2$ is therefore a function of the momentum distribution
of the slepton, which is determined by the details of the SUSY spectrum.
One needs therefore to be able to simulate the full SUSY cascade decays
in order to construct the $c\tau$--$R$ relationship. 

\begin{figure} 
\epsfxsize = \textwidth
\epsffile{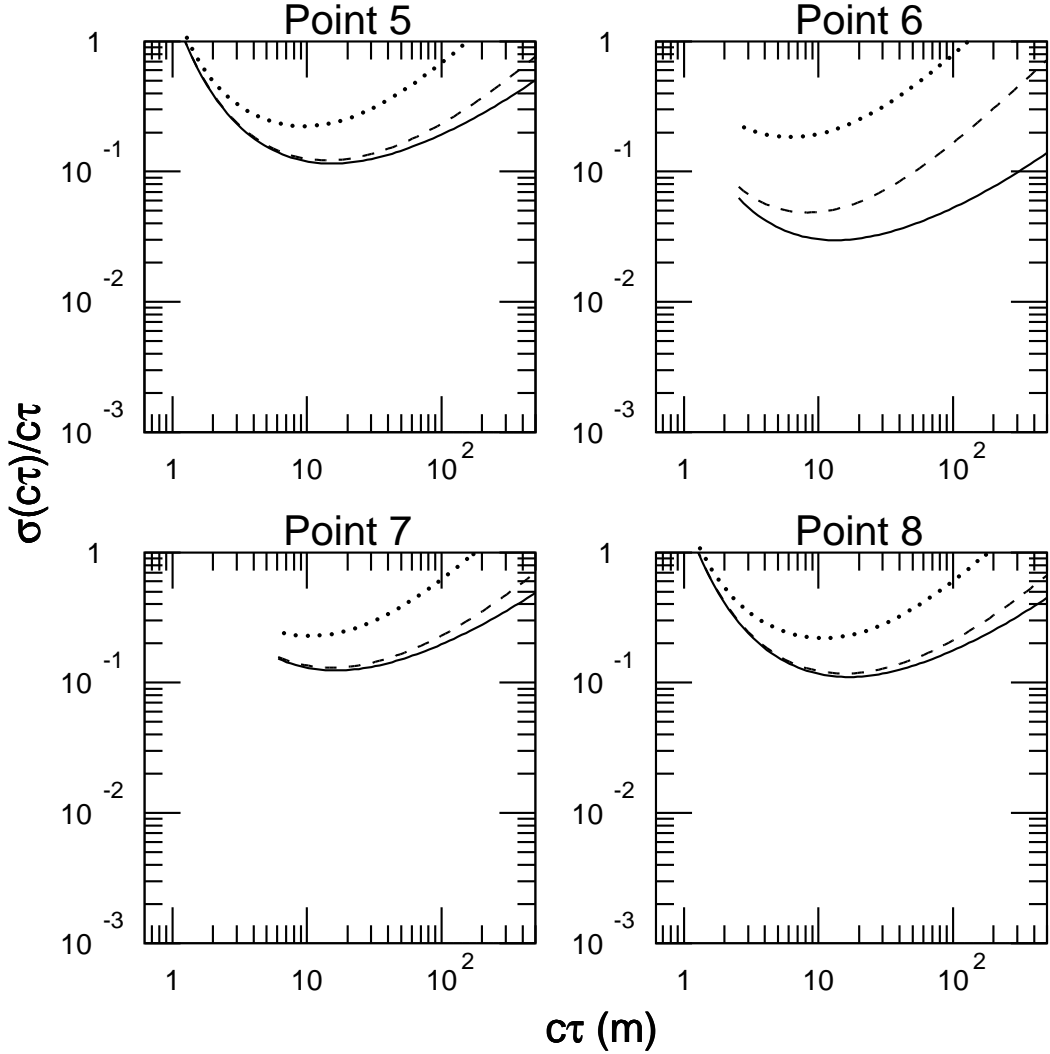}
\caption{\sl The same as in Fig.~\ref{fig:sigctau2}, but for 
the model sample points B5 to B8.
}
\label{fig:sigctau2}
\end{figure}

The statistical error on $R$ can be evaluated as

\be
\sigma(R)=\sqrt{\frac{R(1-R)}{N_1}}.
\label{eq:errat} 
\ee

Relevant for the precision with which the SUSY breaking scale can be 
measured is instead the error on the measured $c\tau$. This can be 
extracted from the curves shown in Fig.~\ref{fig:ratio} and can be 
evaluated as

\be
\sigma({\mathrm c}\tau)=\sigma(R)/
\left[\frac{\partial R(c\tau)}{\partial c\tau}\right].
\label{eq:erctau}
\ee

The measurement precision calculated according to this formula is shown 
in Figs.~\ref{fig:sigctau1} and \ref{fig:sigctau2} for the eight sample 
points, for an integrated luminosity of 30~fb$^{-1}$. 
The full line in the plots is the error on $c\tau$ considering the statistical
error on $R$ only. The available statistics is a function 
of the strongly interacting sparticles' mass scale. 
Even if a precise $R$--$c\tau$ relation can be built from the 
knowledge of the model details, there will be a systematic uncertainty 
in the evaluation of the losses in $N_2$, because of sleptons produced outside
the $\eta$ acceptance, or absorbed in the calorimeters, or escaping the 
calorimeter with a transverse momentum below the cuts. 
The full study of these uncertainties is in progress. 

\begin{figure}[ht]
\epsfxsize = \textwidth
\epsffile{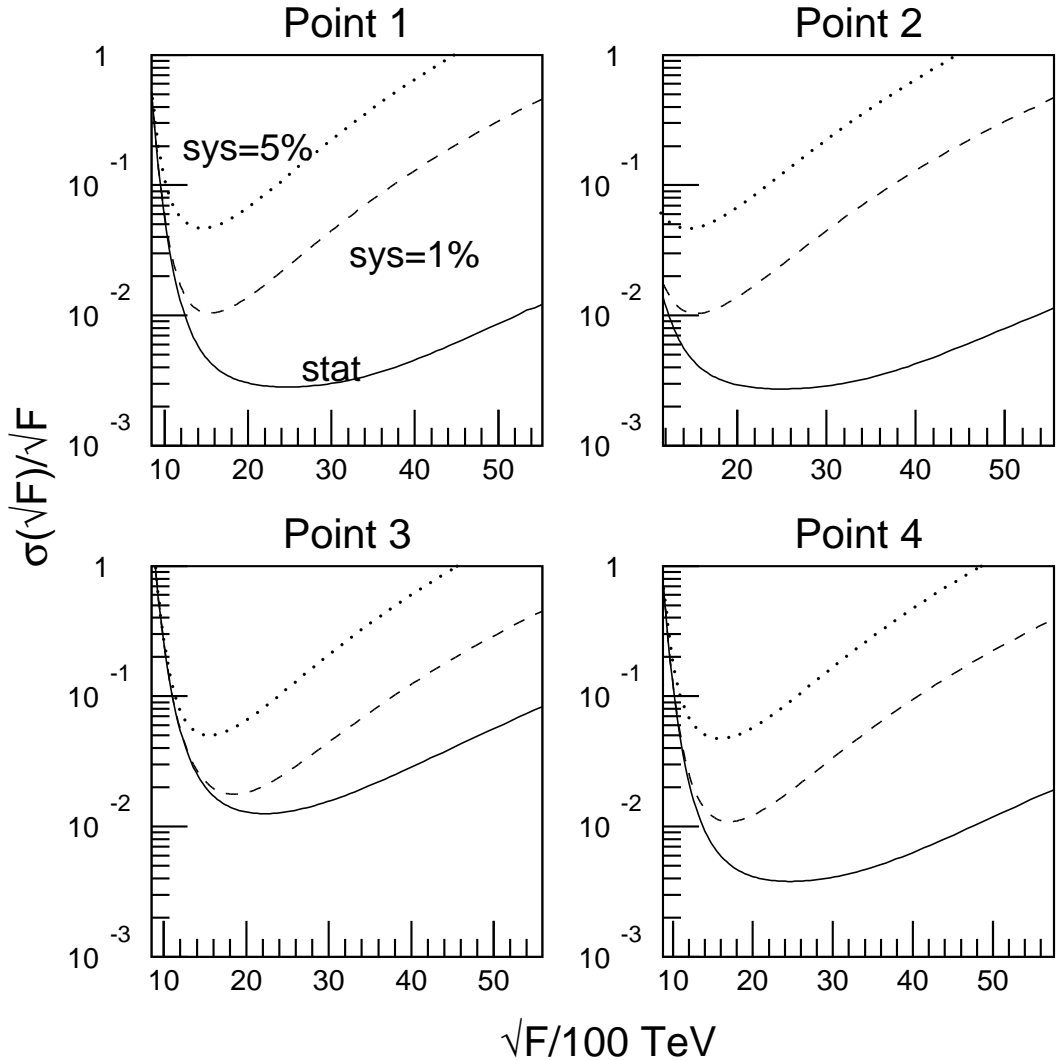}
\caption{\sl Fractional error on the measurement of the 
SUSY breaking scale $\sqrt{F}$ for model sample points B1 to B4.
We assume an integrated luminosity of 30~fb$^{-1}$. 
The curves are shown for the three different assumptions on the fractional 
systematic error used in Figs.~\ref{fig:sigctau1} and \ref{fig:sigctau2}.
}
\label{fig:sigsqrtf1}
\end{figure}

At this level, we just parameterise the systematic error as a term 
proportional to $R$, added in quadrature to the statistical error. 
We choose two values, $1\% R$ and $5\% R$, and propagate the error to 
the $c\tau$ measurement. 
The results are represented by the dashed and dotted lines in
Figs.~\ref{fig:sigctau1} and \ref{fig:sigctau2}. 

For the models with squark mass scales up to 1200~GeV,
assuming a 1\% systematic error on the measured ratio, 
a precision better than 10\% on the $c\tau$ measurement 
can be obtained for lifetimes between 0.5--1~m and 50--80~m. 
If the systematic uncertainty grows to 5\%, the 10\% precision
can only be achieved in the range 1--10~m. If the mass scale goes up to 2~TeV,
even considering a pure statistical error only, a 10\% precision
is not achievable. However a 20\% precision is possible  
over $c\tau$ ranges between 5 and 100~m, assuming a 1\% systematic error.

\begin{figure}
\epsfxsize = \textwidth
\epsffile{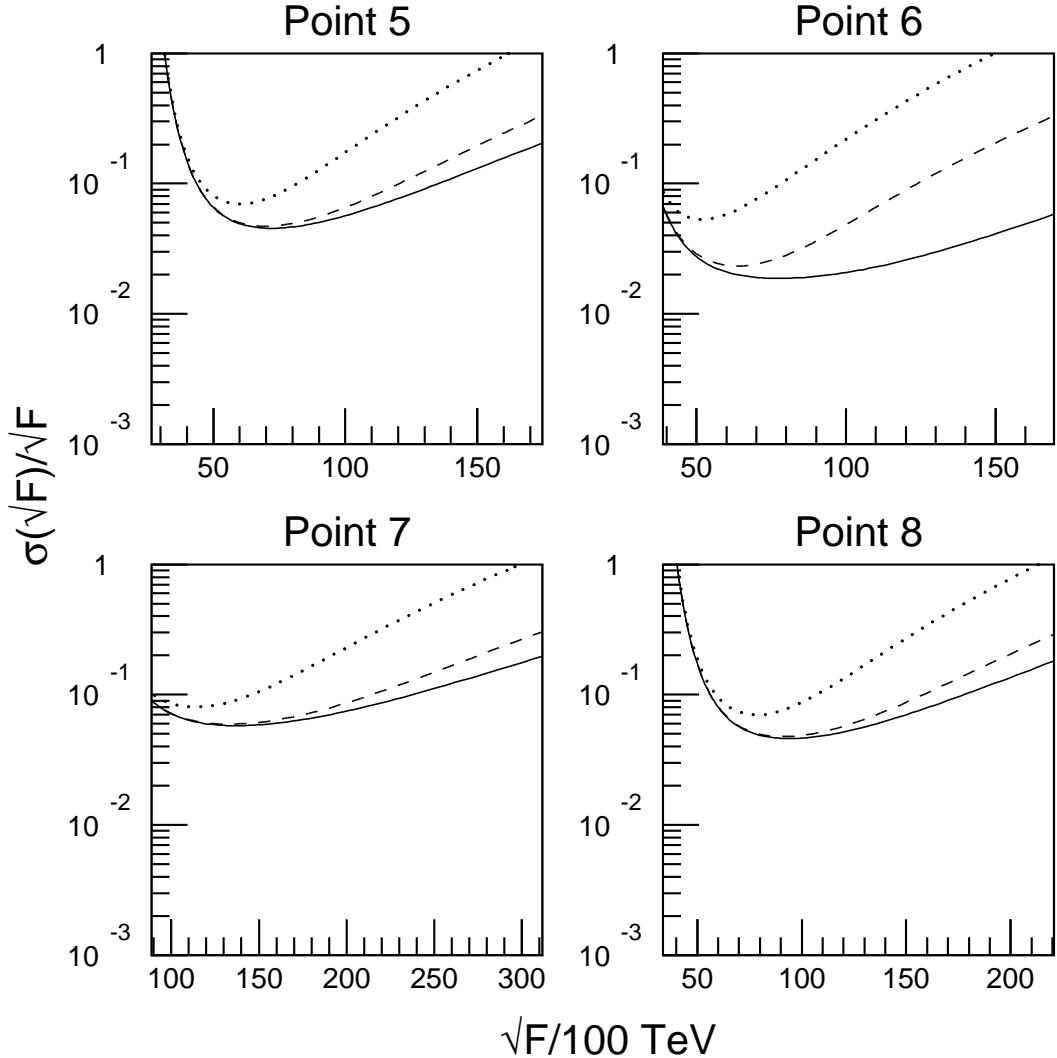}
\caption{\sl The same as in Fig.~\ref{fig:sigsqrtf1}, but for model
sample points B5 to B8.
}
\label{fig:sigsqrtf2}
\end{figure}

Note that the curves corresponding to the model points B2, B6 and B7 
do not start from $c\tau = 50$ cm, but from the theoretical lower 
limit on $c\tau$ of 1.8, 2.5 and 6.1 metres, respectively.

\section*{B.6~~Determining the SUSY Breaking Scale $\sqrt{F}$}
\label{sec:B6}

\noindent 
Using the measured values of $c\tau$ and the NLSP mass,
the SUSY breaking scale $\sqrt{F}$ can be calculated from 
Eq.~(\ref{eq:NLSPtau}), where ${\cal B} = 1$ for the case where 
the NLSP is a slepton.  
From simple error propagation, the fractional uncertainty on the 
$\sqrt{F}$ measurement can be obtained adding in quadrature one 
fourth of the fractional error in $c\tau$ and five fourths of the 
fractional error on the slepton mass. 

In Figs.~\ref{fig:sigsqrtf1} and \ref{fig:sigsqrtf2}, we show the fractional 
error on the $\sqrt{F}$ measurement as a function of $\sqrt{F}$ for 
our three different assumptions on the $c\tau$ error. 
The uncertainty is dominated by $c\tau$ for the higher part of the 
$\sqrt{F}$ range and grows quickly when approaching
the lower limit on $\sqrt{F}$. This is because very few sleptons 
survive and the statistical error on both $m_{\tilde \ell}$ and 
$c\tau$ gets very large. 
If we assume a 1\% systematic error on the ratio $R$ from which $c\tau$
is measured (dashed lines in Figs.~\ref{fig:sigsqrtf1} and 
\ref{fig:sigsqrtf2}), the error on $\sqrt{F}$ is better than 10\% 
for $1000 \ltap \sqrt{F} \ltap 4000$~TeV for model points B1--B4 with
higher statistics. For points B5--B8, in general one can explore 
a range of higher $\sqrt{F}$ values with a small relative error, 
essentially due to the heaviness of the decaying NLSP in these models.
Note also that the theoretical lower limit (\ref{eq:sqrtFmin}) on $\sqrt{F}$ 
is equal to about 1200, 1500, 3900, 8900 TeV respectively in model 
points B2, B5, B6, B7, while it stays well below 1000 TeV for the other models.

\section*{B.7~~Conclusions}
\label{sec:B7}

\noindent 
We have discussed a simple method to measure at the LHC with the ATLAS 
detector the fundamental SUSY breaking scale $\sqrt{F}$ in the GMSB scenarios 
where a slepton is the NLSP and decays to the gravitino with a lifetime
in the range 0.5~m $\ltap c\tau_{\rm NLSP} \ltap 1$~km. This method requires
the measurement of the time of flight of long lived sleptons 
and is based on counting events with one or two identified NLSP's.  
It relies on the assumptions that a good knowledge of the MSSM sparticle 
spectrum and BR's can be extracted from the observation of the SUSY events 
and that the systematic error in evaluating the slepton losses can be kept
below the few percent level.
We performed detailed, particle level simulations for eight representative 
GMSB models, some of them being particularly hard due to low statistics. 
We found that a level of precision of a few 10's \% on the SUSY breaking 
scale measurement can be achieved in significant parts of the 
$1000 \ltap \sqrt{F} \ltap 30000$~TeV range, for all models considered. 
More details as well as a full study of the systematics associated with this 
procedure and another less ``model-dependent'' method to measure $\sqrt{F}$ 
is presented in detail in Ref.~\cite{AMPPR}. 

\bigskip
\subsection*{Acknowledgements}
S.~A. and G.~P. thank the organisers of the Workshop ``Physics at TeV
Colliders'', for the hospitality and the pleasant and productive atmosphere
in Les Houches.

\end{document}

%% file: SA_macros.tex
\newcommand{ \be  }{\begin{equation}}
\newcommand{ \ee  }{\end{equation}} 
\newcommand{ \bea }{\begin{eqnarray}}
\newcommand{ \eea }{\end{eqnarray}}
\newcommand{ \beas }{\begin{eqnarray*}}
\newcommand{ \eeas }{\end{eqnarray*}}

\newcommand{ \NPB    }[3]{Nucl. Phys.      {\bf B#1} (#2) #3}
\newcommand{ \PLB    }[3]{Phys. Lett. B    {\bf #1}  (#2) #3}
\newcommand{ \PLBold }[3]{Phys. Lett.      {\bf #1B} (#2) #3}
\newcommand{ \PRD    }[3]{Phys. Rev. D     {\bf #1}  (#2) #3}

\newcommand{ \PRL    }[3]{Phys. Rev. Lett. {\bf #1}  (#2) #3}

\newcommand{ \ZPC    }[3]{Z. Phys. C       {\bf #1}  (#2) #3}
\newcommand{ \EPJC   }[3]{Eur. Phys. J. C  {\bf #1}  (#2) #3}

\newcommand{ \centeron }[2]{{\setbox0=\hbox{#1}\setbox1=\hbox{#2}\ifdim
                             \wd1>\wd0\kern.5\wd1\kern-.5\wd0\fi \copy0
                             \kern-.5\wd0\kern-.5\wd1\copy1\ifdim\wd0>\wd1
                             \kern.5\wd0\kern-.5\wd1\fi}}
\newcommand{ \ltap }{\;\centeron{\raise.35ex\hbox{$<$}}
                     {\lower.65ex\hbox{$\sim$}}\;}
\newcommand{ \gtap }{\;\centeron{\raise.35ex\hbox{$>$}}
                     {\lower.65ex\hbox{$\sim$}}\;}

\newcommand{ \slashchar }[1]{\setbox0=\hbox{$#1$}   
   \dimen0=\wd0                                     
   \setbox1=\hbox{/} \dimen1=\wd1                   
   \ifdim\dimen0>\dimen1                            
      \rlap{\hbox to \dimen0{\hfil/\hfil}}          
      #1                                            
   \else                                            
      \rlap{\hbox to \dimen1{\hfil$#1$\hfil}}       
      /                                             
   \fi}                                             %

\newcommand{ \NI         }{ \tilde{N}_1 }

\def\G{ \tilde{G} }

\newcommand{ \lR         }{ { \tilde \ell}_R }

\newcommand{ \tauu       }{ { \tilde \tau}_1 }

\newcommand{\sss}{\scriptscriptstyle}

\newcommand{\Z}{Z^{\sss 0}}

\newcommand{\epem}{e^{\sss +}e^{\sss -}}

\newcommand{\stopu}{\tilde{t}_1}
\newcommand{\stopd}{\tilde{t}_2}

